# Monetary Policy and Economic Growth in Developing Countries: A Literature Review

**Marouane Daoui**

Ph.D. in Economics and Management, Faculty of Law, Economics and Social Sciences, Sidi Mohamed Ben Abdellah University, Fez, Morocco, marouane.daoui@usmba.ac.ma.

**Abstract**

This article conducts a literature review on the topic of monetary policy in developing countries and focuses on the effectiveness of monetary policy in promoting economic growth and the relationship between monetary policy and economic growth. The literature review finds that the activities of central banks in developing countries are often overlooked by economic models, but recent studies have shown that there are many factors that can affect the effectiveness of monetary policy in these countries. These factors include the profitability of central banks and monetary unions, the independence of central banks in their operations, and lags, rigidities, and disequilibrium analysis. The literature review also finds that studies on the topic have produced mixed results, with some studies finding that monetary policy has a limited or non-existent impact on economic growth and others finding that it plays a crucial role. The article aims to provide a comprehensive understanding of the current state of research in this field and to identify areas for future study.

## 1. Introduction

Monetary policy in developing countries is a topic that has recently gained more attention among economists. Economic models tend to focus on the activities of central banks in developed countries, while the activities of central banks in developing countries are often overlooked. However, recent studies have shown that there are many factors that can affect the effectiveness of monetary policy in these countries, including the profitability of central banks and monetary unions. Central banks in developing countries may also lack independence in their operations, which can lead to unclear objectives for monetary policy. In this context, researchers have argued that lags, rigidities, and disequilibrium analysis are key factors in understanding short-term macroeconomics in these countries. This introduction provides an overview of the current state of research on monetary policy in developing countries and highlights some of the key issues and challenges that central banks in these countries face.

Monetary policy refers to the actions taken by a central bank to control the money supply and interest rates in an economy. According to economists such as Friedman, Cagan, Mishkin, Tanzi, and others, an increase in the money supply leads to an increase in the availability of credit and a reduction in the interest rate is needed to stimulate the demand for money. However, the effectiveness of monetary policy is influenced by various factors such as the elasticity of money demand with respect to income and interest rates, the elasticity of aggregate spending with respect to interest rates, and the mobility of capital in an open economy. Furthermore, factors on the supply side of the economy such as labor market conditions and capacity constraints can also impact the effectiveness of monetary policy. Additionally, the objectives of monetary policy and the government's budget deficit can also play a role in determining the impact of monetary policy on the economy. Overall, a comprehensive analysis of the effectiveness of monetary policy requires an assessment of the flexibility and constraints on both the supply and demand sides of the economy.

The relationship between monetary policy and economic growth has been the subject of much research, with varying results. Some studies have found that monetary policy has a limited or non-existent impact on economic growth, while others have found that it plays a crucial role. Studies such as Cyrus and Elias (2014) and Lashkary and Kashani (2011) have found that monetary policy has a negligible influence on real output, while others such as Havi and Enu (2014) and Rasaki et al. (2013) have found that monetary policy has a significant positive impact on economic growth. The literature also shows that the impact of monetary policy can vary depending on the country and the specific monetary policy measures used.

This paper aims to provide an overview of the current research on the topic of monetary policy in developing countries. Specifically, it will cover the specificity of monetary policy in these countries, the effectiveness of monetary policy in promoting economic growth, and a survey of the empirical literature on the relationship between monetary policy and economic growth in developing countries. The goal is to provide a comprehensive understanding of the current state of research in this field and to identify areas for future study.

## 2. Specificity of Monetary Policy in Developing Countries

Economic models that attempt to identify a reaction function for the monetary authority focus on the activities of central banks in developed countries. In contrast, the activities of central banks in developing countries do not receive the same attention, as it is assumed that central banks in these countries were created with the primary objective of financing the public deficit. However, there is an increasing interest in studying monetary policy in developing countries. Recent studies have looked at various factors that may affect the effectiveness of monetary policy, such as the profitability of central banks (Buiter, 2008) and monetary unions (Kamar and Naceur, 2007). Indeed, the authors have distinguished between two types of monetary policy: accommodative and stabilizing. An accommodative monetary policy is defined as a policy that provides a steady supply of credit to an expanding economy. A stabilizing monetary policy, on the other hand, is a policy that serves to mitigate or offset unwanted changes in the economy. In the first accommodative scenario, money growth accommodates output growth and price inflation. In the second stabilizing scenario, the monetary authority varies money growth to counter the effects of other shocks, depending on the objectives of monetary policy.

Many central banks in developing countries lack independence in their operations, resulting in unclear objectives for monetary policy. Researchers such as Behrman (1981) and Crockett (1981) have argued that lags, rigidities, and disequilibrium analysis are key factors in understanding short-term macroeconomics in these countries. Porter and Ranney (1982) have also focused on analyzing the structural differences between developing and developed countries. The stabilizing function may be to target economic growth. Thus, the central bank increases liquidity to enhance credit expansion and stimulate the economy during a downturn. In an extreme scenario, the government could play a leading role in structural, economic and social development. Indeed, the central bank may be forced to extend credit to finance increased government spending.

However, a problem arises when increased government spending limits available credit and crowds out private spending (Bean and Buiter, 1989). Some studies have focused on specific factors that determine the crowding-out effect of government spending. One factor is the impact of government spending on interest rates (Evans, 1987). Another strand of literature focuses on the sensitivity of investment demand to changes in interest rates (Chirinko, 1993). Moreover, increased spending cannot be properly directed toward expanding productive capacity. As a result, higher government spending can be inflationary. Financing increased government spending through monetization depletes the stock of foreign exchange reserves in a small open economy. As a result, devaluation of the domestic economy may be necessary, increasing the risk of a vicious cycle of depreciation and inflation.

Given concerns about rising government spending, central banks may use nominal anchors to guide the design of the monetary policy. In a small open economy, priorities can be set to defend an exchange rate anchor. As a result, monetary policy independence is severely compromised, as interest rate movements can contradict the exchange rate target and the level of foreign exchange reserve support.

Some developing countries, however, may decide that the exchange rate anchor is not the optimal anchor for monetary policy if they suffer from high inflation. Therefore, inflation targeting may be a better nominal anchor for monetary policy design. Indeed, monetary authorities establish a variety of indicators that guide the inflation process. The money supply responds to changes in these indicators to control inflation. This scenario removes a great deal of independence from the monetary authority.

In a scenario that involves a higher degree of independence, the central bank may take a more discretionary approach. As a result, priorities may be set, and not necessarily announced, in response to economic development. The design of monetary policy may aim at exchange rate stability, price level stability or output growth.

Whatever the objectives set for monetary policy, the ultimate result of money supply fluctuations will be absorbed by economic growth and price inflation. The transmission mechanism of monetary policy to the real economy is aggregate demand. The magnitude of the shift in aggregate demand depends on the liquidity effect attributed to a change in the money supply and the sensitivity of aggregate demand to that change. However, crowding out can occur if inflationary expectations rise and/or capital outflows increase.

Assuming that monetary policy is effective in stimulating demand growth, the allocation of the change in demand between output growth and price inflation depends on the supply-side constraints of the economy. Capacity constraints are bound to accelerate price inflation. Nevertheless, wage and/or price rigidity can reinforce the real effect and moderate the nominal effects of money supply fluctuations.

### 3. Effectiveness of Monetary Policy

Theoretically, an increase in the money supply leads to an increase in the availability of credit. This concept is supported by economists such as Friedman (1968), Cagan (1972), Mishkin (1981), Tanzi (1984), and Mishkin (1988). To return the money market to equilibrium, a reduction in the interest rate is needed to stimulate the demand for money. To restore equilibrium in the money market, a reduction in interest rates is necessary to stimulate the demand for money. This has been studied by economists such as Colleman, Gilles, and Labadie (1992), Leeper and Gordon (1992), Fama (1990), Mishkin (1992), Wallace and Warner (1993), Evans and Lewis (1995), and Soderlind (1997, 1999) who have examined the liquidity effect of money growth. In the goods market, aggregate spending increases in response to a decrease in the interest rate. Higher income stimulates the demand for money, thus moderating the necessary reduction in the interest rate. The effectiveness of monetary policy in stimulating aggregate demand thus depends on three factors: (i) the elasticity of money demand with respect to a change in income, (ii) the elasticity of money demand with respect to a change in the interest rate, and (iii) the elasticity of aggregate spending with respect to a change in the interest rate.

The effectiveness of monetary policy decreases as the response of money demand to income increases. After an increase in the money supply, a stronger money demand response to an increase in income closes the disequilibrium gap and reduces the required reduction in the interest rate. The efficiency of monetary policy makes the demand for money less sensitive to a change in the interest rate. Therefore, a large reduction in the interest rate is necessary in the face of an increase in the money supply. The effectiveness of monetary policy also increases the sensitivity of aggregate demand to a change in the interest rate. As a result, the magnitude of changes in aggregate demand is maximized in the event of monetary shocks.

The above scenario is valid in a closed economy. In an open economy, changes in interest rates can affect the mobility of capital. A reduction in the interest rate induces capital outflows and reduces capital inflows. When reserves are reduced, the monetary base also decreases, limiting the impact of monetary policy. Exchange rate movements can also play a role in this context. When international reserves decrease, monetary authorities may choose to slow domestic credit growth in order to increase domestic interest rates and attract foreign funds. Under a floating exchange rate, an increase in net capital outflows leads to a depreciation of the exchange rate, which increases trade competitiveness. The resulting increase in net exports reinforces the expansionary effect of monetary policy.

The impact of monetary policy is also influenced by factors on the supply side of the economy. For instance, Lane and Perotti (1996) examined labor market conditions to evaluate supply limitations. The more flexible wages and prices are, the greater the inflationary effect of monetary policy and the smaller the expansion of output. Capacity constraints may also limit the response of output to monetary expansion, requiring a more rapid adjustment in price inflation. Faster price inflation reduces real money balances and raises the interest rate, offsetting the positive effect of monetary policy on aggregate demand. The magnitude of the price increase determines the crowding out effect attributed to product market conditions.

Other factors may further complicate the analysis of the effectiveness of monetary policy. The objectives of monetary policy are also of some relevance. If monetary growth is accompanied by an increase in the government's budget deficit, the central bank may deplete its limited stock of foreign exchange reserves to finance increased government spending. In addition, it is likely that inflationary expectations will build up, leading to a decline in the demand for money and an acceleration of price inflation. When confidence in the stability of the exchange rate is lost, devaluation becomes unavoidable. This can further fuel inflationary expectations and weaken the impact of monetary policy (Caballero and Pyndick, 1996). In an open economy, uncertainty can deter foreign direct investment and prompt capital outflows.

Given the channels of interaction in the macroeconomy, a study of the effectiveness of monetary policy requires an assessment of the parameters that measure flexibility and identify constraints on the supply and demand sides of the economy.

## 4. A Survey of the Empirical Literature on the Relationship between Monetary Policy and Economic Growth

Extensive work has been done to try to establish the impact of monetary policy on economic growth, but without much consensus so far. Some studies have confirmed that the impact of monetary policy is limited or non-existent. Cyrus and Elias (2014), applying recursive VAR methodology on time series data from 1997 to 2010, estimated the impact of monetary and fiscal policy shocks on economic growth in Kenya. They find that monetary policy (both money supply and short-term interest rates) has a negligible influence on real output. They argue that the weak link is attributed to the weak structural, institutional, and regulatory framework (Cyrus and Elias, 2014). Using econometric regression model analysis on a monetarist approach, Lashkary and Kashani (2011) studied the impact of monetary variables on economic growth in Iran during the period 1959-2008 and found no significant relationship between money supply and real economic variables, economic growth, and employment (Lashkary and Kashani,2011).

However, several empirical studies confirm that monetary policy is crucial for economic growth. Havi and Enu (2014) examine the relative importance of monetary policy and fiscal policy on economic growth in Ghana over the period 1980 to 2012. The results of ordinary least squares (OLS) estimation revealed that monetary policy as a measure of money supply had a significant positive impact on the Ghanaian economy (Havi and Enu, 2014). Vinayagathasan (2013) estimates the impact of monetary policy on the real economy using a seven-variable structural VAR model using monthly time series data from Sri Lanka covering the period from January 1978 to December 2011. The study finds that interest rate shocks have a significant impact on output, consistent with economic theory. It also finds that a positive currency shock yields significant but inconsistent results on output. Output decreases rather than increases (Vinayagathasan, 2013).

Rasaki et al. (2013) used the OLS method and correlation matrix to examine the impact of fiscal and monetary policies on economic growth in Nigeria, with particular reference to the period between 1998 and 2008. They found that the monetary variables of narrow money and broad money are important policy variables that have a positive effect on economic growth (real GDP growth rate) in Nigeria (Rasaki, Afolabi, Raheem and Bashir 2013).

Davoodi, Dixit, and Pinter (2013) used three variants of structural VAR models on monthly databases from 2000 to 2010 to determine the monetary transmission mechanisms in the East African Community. The study found that the monetary transmission mechanism tends to be generally weak when using standard statistical inferences, but somewhat stronger when using non-standard inference methods. Expansionary monetary policy (a positive shock to money reserves) increases output significantly in Burundi, Rwanda, and Uganda. However, they also find that an expansionary monetary policy (a negative shock to the policy rate) increases output in Burundi, Kenya, and Rwanda (Davoodi, Dixit and Pinter 2013). Berg et al. (2013) used the narrative approach pioneered by Romer and Romer (1989) to examine monetary transmission mechanisms by focusing on four East African countries (Uganda, Kenya, Tanzania, and Rwanda). They found clear evidence of an effective transmission mechanism: after a large policy-induced increase in the short-term interest rate, lending and other interest rates rise, the exchange rate tends to appreciate, and output growth tends to fall (Berg, Charry, Portillo and Vlcek 2013).

Fasanya, Onakoya, and Agboluaje (2013) examined the impact of monetary policy on economic growth in Nigeria using the error correction model (ECM) on time series data covering the period 1975 to 2010. They found that a long-run relationship exists between the variables and that the inflation rate, exchange rate and external reserves are important policy instruments that drive growth in Nigeria in line with theoretical expectations. The money supply was found to be insignificant (Fasanya, Onakoya and Agboluaje 2013).

Onyeiwu (2012), who examined the impact of monetary policy on the Nigerian economy using the OLS method to analyze data between 1981 and 2008, found that monetary policy as represented by the money supply has a positive impact on GDP growth.

Coibion (2012) estimated the effects of monetary shocks on the U.S. economy for the period 1970-1996, using the standard VAR model versus the large effects of the Romer and Romer (2004) approach. The study found that with the standard VAR model approach, monetary policy shocks appear to account for a very small share of fluctuations in the real economy, as measured by either industrial output or unemployment. It was also found that the 1980-1982 and 1990 recessions could not be explained by the standard VAR model (Coibion, 2012). Estimates of a DSGE model by Smets and Wouters (2007) took into account the average size effects of monetary shocks on real variables, including output (Coibion, 2012).

Jawaid, Qadri, and Ali (2011) studied the effect of monetary, fiscal, and trade policy on economic growth in Pakistan, using annual time series data from 1981 to 2009. They used cointegration and error correction model (ECM), revealing the existence of a significant positive relationship in the long and short run between monetary policy (money supply) and economic growth (Jawaid, Qadri and Ali, 2011).

Amarasekara (2009) used the recursive VAR and semi-structural VAR methodology on monthly data for the period 1978 to 2005 to assess the effects of monetary policy on economic growth and inflation in the small developing open economy of Sri Lanka. The results of the recursive VAR model were consistent with the results of the semi-structural VAR model and revealed a significant negative impact of interest rates on growth. Positive innovations reduced GDP growth. However, when money growth and the exchange rate are used as policy indicators, the impact on GDP growth contrasts with the results obtained with the theory (Amarasekara, 2009). Muhammad et al. (2009) used Johnson's cointegration test to study the long-run relationship between the money supply (M2), government spending, and economic growth in Pakistan using annual data for the period 1977-2007. They found a positive relationship between money supply (M2) and economic growth in the long run (Muhammad, Wasti, Lal, and Hussain, 2009).

Rafiq and Mallick (2008) examined the effects of monetary policy shocks on output in three euro-area economies (Germany, France, and Italy) by applying a new VAR model identification procedure. The results showed that monetary policy innovations are most powerful in Germany. However, with the exception of Germany, it remained ambiguous whether a rise in interest rates coincided with a fall in output, showing a lack of homogeneity in the responses. They concluded that monetary policy innovations play a modest role in generating output fluctuations for the European Monetary Union (Rafiq, and Mallick, 2008).

Balogun (2007) examines the monetary and macroeconomic stability perspective of the West African currency area countries using a sample of quarterly data from 1991:Q1 to 2004:Q4. The regression results indicate that monetary policy, as captured by the money supply and credit to government, has been detrimental to real domestic output in these countries. The study also shows that interest rate policy had negative effects on GDP contrary to the theoretical expectation of an inverse relationship and that exchange rate devaluations have no effect on output (Balogun, 2007).

Smets and Wouters (2007) developed and estimated a DSGE model with sticky prices and wages for the euro area. The model was estimated using Bayesian techniques using seven key macroeconomic variables: GDP, consumption, investment, prices, real wages, employment, and the nominal interest rate. In addition, they introduced ten orthogonal structural shocks (including productivity, labor supply, investment, preference, cost-push, and monetary policy shocks) that allowed for an empirical study of the effects of these shocks and their contribution to business cycle fluctuations in the euro area. They found that monetary policy shocks play an important role in changes in euro area output (Smets and Wouters, 2007).

Khabo and Harmse (2014) estimated the impact of monetary policy on South Africa, using OLS on annual data series from 1960 to 1997 and found that money supply (M3) and inflation were significantly related to economic growth, consistent with economic theory (Khabo and Harmse, 2014).

In a recent study, Daoui and Benyacoub (2021a) used a Factor-Augmented Vector Autoregression (FAVAR) approach to analyze the impact of monetary policy shocks on economic growth in Morocco. The study utilizes a large number of Moroccan macroeconomic time series spanning from 1985 to 2018. The FAVAR model, based on dynamic factor models, allows for the summarization of a large database into a small number of factors common to all variables. The results of the analysis indicate that monetary policy shocks have a negative impact on economic growth in Morocco, as evidenced by the overall decline in GDP in response to these shocks.

In addition, Daoui and Benyacoub (2021b) use a more recent extension of the FAVAR model, the Factor-augmented Error Correction Model (FECM), to examine the effects of monetary policy shocks on economic growth in Morocco. The study is based on a database of 117 quarterly series, spanning from 1985 to 2018, and combines the benefits of dynamic factor models and error correction models. The main objective of the study is to explore how monetary policy shocks affect economic growth in Morocco, and to compare the results obtained by the FECM model to those obtained by the factor-augmented vector autoregression (FAVAR) model. The results suggest that the FECM model, which takes into account non-stationarity in dynamic factor modeling, is a valuable extension of the FAVAR model for studying monetary policy shocks.

# 5. Conclusion

The literature on monetary policy in developing countries has traditionally focused on the financing of public deficits by central banks. However, there is an increasing interest in studying the effectiveness of monetary policy in these countries, and recent studies have looked at various factors such as profitability of central banks and monetary unions that may affect this effectiveness. Research suggests that central banks in developing countries may lack independence in their operations, resulting in unclear objectives for monetary policy. Some studies have focused on specific factors that determine the impact of government spending on interest rates and investment demand. As a result, central banks in developing countries may use nominal anchors such as exchange rate or inflation targeting to guide monetary policy, but this may compromise monetary policy independence. Ultimately, the success of monetary policy in stimulating economic growth and controlling inflation depends on the supply-side constraints of the economy, and the risk of crowding out private spending by government spending.

The effectiveness of monetary policy in stimulating aggregate demand depends on several factors, including the elasticity of money demand with respect to changes in income and interest rates, and the elasticity of aggregate spending with respect to changes in interest rates. The demand for money becomes less sensitive to changes in interest rates as the efficiency of monetary policy increases. In an open economy, changes in interest rates can also affect the mobility of capital and exchange rates, further complicating the analysis of the effectiveness of monetary policy. Additionally, other factors such as labor market conditions, capacity constraints, and government budget deficits can also influence the impact of monetary policy. Overall, a comprehensive assessment of the various parameters that measure flexibility and identify constraints on the supply and demand sides of the economy is necessary to study the effectiveness of monetary policy.

The impact of monetary policy on economic growth in developing countries has been studied extensively, but there is still no consensus on the subject. Some studies have found that the impact of monetary policy is limited or non-existent. On the other hand, several empirical studies confirm that monetary policy is crucial for economic growth. Overall, the studies reviewed suggest that the relationship between monetary policy and economic growth is complex and varies by country, and further research is needed to better understand the relationship.